\documentstyle[12pt]{article}

\newcommand{\be}{\begin{equation}}
\newcommand{\ee}{\end{equation}}
\newcommand{\beqs}{\begin{eqnarray}}
\newcommand{\eeqs}{\end{eqnarray}}

\begin{document}

\pagestyle{plain} \setcounter{page}{1} \newcounter{bean} \baselineskip16pt


\begin{titlepage}
\begin{flushright}
PUPT-1743\\

\end{flushright}

\vspace{7 mm}

\begin{center}
{\huge String Theory and Quark Confinement}

\vspace{5mm}

\end{center}
\vspace{10 mm}
\begin{center}
{\large
Alexandre M.~Polyakov\\
}
\vspace{3mm}
Joseph Henry Laboratories\\
Princeton University\\
Princeton, New Jersey 08544\\
\vspace{3mm}
\end{center}
\vspace{7mm}
\begin{center}
{\large Abstract}
\end{center}
\noindent This article is based on a talk given at the 
``Strings '97'' conference. 
It discusses the search for the universality class of confining strings. 
The key ingredients include 
the loop equations, the zigzag symmetry, 
the non-linear renormalization group. 
Some new  tests for the equivalence between 
gauge fields and strings are proposed.  
\vspace{7mm}
\begin{flushleft}
October 1997

\end{flushleft}
\end{titlepage}

This talk is a brief summary of recent results on the string representation
of gauge theories (or gauge theory representation of strings).
This subject
is twenty years old, but the last word on it has not been spoken. In our
opinion this is one of the major challenges of theoretical physics.

The original project (proposed in [1] and developed in [2, 3]) to solve
this problem has taken the following route. As a first step we established
the equation for the Wilson loop which, being the variational equation in
the loop space, encoded the Yang Mills equations in the ordinary space. The
next step should be the solution of the loop equation by the string
functional integral, analogous to the solutions of the wave equations by the
Feynman integrals over trajectories.

In other words, let us consider the Wilson loop as a functional of a contour 
$C$%
\begin{equation}
W[c(s)]=\frac{1}{N}\langle TrP\exp \oint_{C}A_{\mu }dx_{\mu }\rangle
\end{equation}
where $A_{\mu }$ is the gauge field for the $SU(N)$
group and c$_{\mu
}=c_{\mu }(s)$ is a parametrized contour.
The brackets mean the average over
the gauge fields with the Yang Mills action. The Schwinger Dyson equations
for $A_{\mu }$translate into the loop equations (we take the large N limit
for simplicity): 
\begin{equation}
\widehat{L}W(C)=g_{0}^{2}\int dsdu\dot{c}(s)\dot{c}(u)\delta (c(s)-c(u))W(%
\overline{C})W(\overline{\overline{C}})
\end{equation}
where $g_{0}$ is the Yang Mills coupling constant and the loop operator $%
\widehat{L}$ is given by [3] 
\begin{equation}
\widehat{L}=\lim_{\varepsilon \rightarrow 0}\oint ds\int_{-\varepsilon
}^{\varepsilon }dt\frac{\delta ^{2}}{\delta c(s+t)\delta c(s-t)}
\end{equation}

In the string picture we interpret $W[C]$ 
as a sum over surfaces with disk
topology, bounded by the contour $C$, or as an amplitude for the open
string the ends of which trace the contour.

The loop equation tells us that when the ends of this open string are hit
with the operator $\widehat{L}$ we get ''almost'' zero, or more precisely
the contribution of the pinched disk with the boundary formed by $\overline{C%
}$ and $\overline{\overline{C}}$ .

In the standard string theory there are many relations of a similar
nature, although they are explicitly known only in the strings with $c<1$
where they were derived by the use of the matrix models equations of motion
in the remarkable paper [4]. In the general case they appear for the
following reason [5]. Let us consider a functional integral over the world
sheet metrics and examine the expression 
\begin{equation}
\int D\mu (g_{ab})\frac{\delta }{\delta g^{ab}}e^{-S}=\int D\mu
(g_{ab})T_{ab}e^{-S}
\end{equation}

The naive value of this expression is zero while in reality this integral is
dominated by the boundaries in the space of metrics. Therefore we may expect
a new type of anomalies in the Schwinger-Dyson equations (a possibility
noted in [6] ). These boundaries are not empty because we assume that the
metric $g_{ab}$ and hence its eigenvalues $\zeta _{1,2}$ are positive
definite. Therefore we can have two types of degenerations. The first type
which we will call ''pinch'' occurs when $\zeta _{1}=0$ , $\zeta _{2}>0.$
This singularity might seem strange, because in the conformal gauge the
eigenvalues of the metric are always equal. The situation becomes clear if
we consider the following example. Take a metric of the form 
\begin{equation}
ds^{2}=dr^{2}+f^{2}(r)d\alpha ^{2}
\end{equation}
with $f^{2}\sim (r-r_{0})^{2}+\varepsilon ^{2}$ and transform this
metric into conformal gauge 
\begin{equation}
ds^{2}=\rho (R)(dR^{2}+R^{2}d\alpha ^{2})
\end{equation}

It is easy to see that as $\varepsilon \rightarrow 0$ the circle $r=r_{0}$
transforms into the circle $R=R_{0}\rightarrow \infty .$ 
The region ``beyond
the horizon'', $r>r_{0}$, represents another sphere attached by the
infinitely thin throat. Another type of degeneration in which the metric
simply tends to zero will be called shrinking.

A proper understanding of the non-critical strings (which we are still
missing) requires the inclusion of the degenerate metrics described above.
Notice that the standard calculations with the Liouville action assume that
such degenerations are absent. We must add them as separate sectors and thus
the typical random surface looks like a collection of smooth components glued
together at certain points. It has been understood long ago [7] that this
can be viewed as a single smooth surface in the self-consistent background,
because the attachment points can be replaced by the vertex operators of
string theory. The relation (4) as well as similar relations, containing
products of the stress tensors, should generate the non-linear equations for
the correlation functions of the theory. In the case of the minimal models
these products seem to be related to the degenerate states. Even in this
simple case we lack a consistent derivation of these equations in the
Liouville theory. Such a derivation presents an important problem, since it
would clarify the connection between the Liouville theory and the matrix
models. More generally, this non-linear structure may form a foundation of
quantum gravity.

It has been further conjectured [5] that the relation (4) results in the
non-linear renormalization group for the partition function, having the
general form 
\begin{equation}
\beta ^{n}(\lambda )\frac{\partial Z}{\partial \lambda ^{n}}=\Gamma
^{nm}(\lambda )\frac{\partial Z}{\partial \lambda ^{n}}\frac{\partial Z}{%
\partial \lambda ^{m}}+a(\lambda )
\end{equation}
Here $\beta (\lambda )$ is the $\beta $ -function of the theory on the world
sheet, $\lambda $ being the coupling constants; the LHS of this equation is
just a trace of the stress tensor. As we mentioned it can also correspond to
the products of these operators. The quantity $\Gamma ^{nm}(\lambda )$ is a
certain metric on the space of the coupling constants. It, as well as a
polynomial $a(\lambda )$, must be calculated dynamically in each particular
case. The first term in the RHS represents the contribution of the pinched
degenerations while the second term, being a polynomial in $\lambda $, is a
contribution of the collapsed surfaces. Sometimes this quantity could be zero
in the continuum limit. The resulting equations can be interpreted as the
Hamilton -Jacobi equations for the corresponding critical string.
Unfortunately we still lack the technique needed to make this conjecture a
precise statement.

Our basic plan for solving the loop equation is now quite straightforward.
We must find a string theory in which the stress tensor is such that at the
boundary of the disk it reduces to the operator $\widehat{L}$ . If we
succeed, the right hand side of the loop equation will arise from the
degenerate metrics. However, a look at the standard string theory reveals a
serious difficulty. In this case the stress tensor has the form 
\begin{equation}
T_{00}\sim \stackrel{.}{x}^{2}+(x^{^{\prime }})^{2}\Rightarrow -\frac{%
\delta ^{2}}{\delta c^{2}(\sigma )}+(\frac{dc}{d\sigma })^{2}
\end{equation}
where $x=x(\sigma ,\tau )$ is the string coordinate, while $c=c(\sigma )$ is
its boundary value. If we insert $T_{00}$ into the disk amplitude we get the
loop operator very different from $\widehat{L}$ . We would also expect that
the pinch at the right hand side of the equation will be saturated by all
possible open string states , while in the gauge theory only the
vector states
appear.

Both discrepancies are due to the fact that the standard non-critical string
has the wrong symmetry for our purposes [8]. It is symmetric under
diffeomorphisms of the world sheet. Correspondingly we can transform 
\begin{equation}
c(s)\Rightarrow c(\alpha (s))
\end{equation}
with $\frac{d\alpha }{ds}>0.$ However we need more than that. The Wilson
loop $W[c(s)]$ is invariant under $s\Rightarrow \alpha (s)$ with any $\alpha
(s).$ Indeed if $\frac{d\alpha }{ds}$ changes sign, the contour $c(\alpha
(s))$ backtracks. The unitarity of the parallel transport insures
that the Wilson loop remains unchanged under this backtracking.

Hence we need a string theory in which both the open string wave operator
and the open string states are invariant under these extended
reparametrizations, which we will call the ''zigzag symmetry''. This symmetry
selects the vector states. 
Indeed let us compare the vector and the tensor vertex
operators 
\begin{equation}
V_{\mu }(p)=\oint ds\stackrel{.}{x}_{\mu }(s)e^{ipx(s)}
\end{equation}
\begin{equation}
V_{\mu \nu }(p)=\oint dsg_{11}^{-\frac{1}{2}}(s)\stackrel{.}{x}_{\mu }(s)%
\stackrel{.}{x}_{\nu }(s)e^{ipx(s)}
\end{equation}
(where $g_{11}$ is a metric at the boundary). The explicit dependence on $%
g_{11}(s)>0$ destroys the zigzag symmetry. That explains why only the vector
states appeared in the RHS of (2). On the other hand , it is clear that the
standard Virasoro generator (8) also breaks the zigzag symmetry. This
becomes clear when one tries to act with this operator on an invariant
object like the ordered exponential. The regularized result of such an
action will contain $\surd g_{11}$ once again.

At the same time the operator $\widehat{L}$ acts nicely on the ordered
exponentials, keeping their zigzag symmetry intact. In fact, it is the most
general operator with this property and the minimal number of derivatives.
This is a counterpart of the fact that the Yang Mills equations are 
a generic consequence of gauge invariance.

We come to the conclusion that the string theory we need is rather peculiar.
While the closed string should have the full-fledged string spectrum, the
open string must have only vector states which are the edge states for the
disk in consideration. The hint of how to achieve it is given again by the
zigzag symmetry. Let us consider a standard string action in a fixed
background 
\begin{equation}
S=\int d^{2}\xi [\surd gg^{ab}(\xi )G_{\mu \nu }(x(\xi ))+\epsilon
^{ab}B_{\mu \nu }(x(\xi ))]\partial _{a}x^{\mu }\partial _{b}x^{\nu
}+...+\oint_{\partial M}B_{\mu }dx^{\mu }
\end{equation}

The zigzag symmetry forbids the first term in this expression since the
presence of the square root requires the positivity of the Jacobian.
Therefore we expect that this term must be absent and the correct
representation for the Wilson loop must be something like 
\begin{equation}
W[C]=\sum_{(S_{C})}\exp (i\int_{S_{C}}B_{\mu \nu }(x(\xi ),\{c(s)\})d\sigma
^{\mu \nu }
\end{equation}
with the background field $B$ which can depend on the contour and must be
determined from the self-consistency.

It has been shown before [8] that in the abelian theory of confinement the
representations of this type arise quite naturally. Summation over surfaces
in this case enforces the Bianchi identity which includes monopoles leading
to confinement. We will discuss possible generalizations of this fact below.

Before coming to the precise definition of the meaning of the sum in this
formula, let us discuss its qualitative significance. The major fact
encoded in it is that the open strings have zero tension. This is seen from
the fact that the string tension is determined by the scale of $G_{\mu \nu }$
which we set to zero. Another useful point of view comes from the T-duality
(the transformation which in the old days was called the Fourier transform).
The loop equation in the momentum representation has been written long ago
in [9]. One defines 
\begin{equation}
\widetilde{W}[p(s)]=\int Dx(s)W[x(s)]e^{i\int pdx}
\end{equation}

It has been noticed in this work that $\widetilde{W}[p(s)]$ is non-zero only
for the piecewise constant $p(s)$ presentable in the form 
\begin{equation}
p(s)=\sum_{k=1}^{L}p_{k}\theta (s-s_{k})
\end{equation}
and that the action of the operator $\widehat{L}$ is given by 
\begin{equation}
\widehat{L}\widetilde{W}=\sum_{k}(p_{k}-p_{k+1})^{2}\widetilde{W}
\end{equation}
as can be easily seen from its definition (2). To interpret this relation let
us consider a collection of D-instantons located at the positions $\{p_{k}\}$
and connected by the open strings going from $p_{k}$ to $p_{k+1}$ .The
Virasoro generator, $L_{0}$ , for these strings has the form 
\begin{equation}
L_{0}\sim \sum_{k}(p_{k}-p_{k+1})^{2}+b^{2}\sum_{n,k}a_{n,k}^{+}a_{n,k}
\end{equation}
where $b^{2}$ is the string tension for the dual string. We see that in the
limit $b\rightarrow \infty $ the excitations of the connecting strings have
infinite energy and hence the Virasoro generator acting on the ground states
of the connecting strings becomes identical with the loop Laplacian. As we
return to the original strings , we obtain the tension $a^{2}\sim
b^{-2}\rightarrow 0$ and we recover the tensionless strings once again.

Is it possible to have a string theory with tensionless open strings and
tensile closed strings? The affirmative answer to this question is based on
the following observation. So far we considered the non-critical strings
described by the coordinates $x^{\mu }(\xi )$ and the world sheet metric $%
g_{ab}(\xi )=e^{\varphi }\delta _{ab}$ where $\varphi (\xi )$ is the
Liouville field. It is well known that non-critical strings in $n$
dimensions can be viewed as the critical strings in the $n+1$ -dimensional
space, formed by $y^{M}=(x^{\mu },\varphi )$. In this description the
background fields $G_{MN}(y),B_{MN}(y),\Phi $ $(y)...$ must satisfy the $%
\beta $ -function equations ( see [10] for a review) which enforce conformal
symmetry on the world sheet. At the same time in the non-critical
description the background fields $G_{\mu \nu }(x),...$ can be chosen
arbitrarily. The relation between these approaches is that the latter fields
fix the boundary values at $\varphi =0$ for the former ones 
\begin{equation}
G_{\mu \nu }(x,\varphi =0)=G_{\mu \nu }(x);...
\end{equation}

It has been conjectured in [5] that the non-linear renormalization group (7)
arises in the critical description as the Hamilton-Jacobi equations satisfied
by the $n+1$ -dimensional effective action as a functional of the boundary
data. An important unsolved problem is to establish these equivalences from
the first principles. Among other things this would elucidate the still
mysterious relation between the continuous 2d quantum gravity and the matrix
models.

We are now in a position to give a concrete meaning to the ansatz (13). The
zigzag symmetry means that we must be looking for the backgrounds satisfying
the conditions 
\begin{equation}
G_{\mu \nu }(x,\varphi =0)=0
\end{equation}

The contour defining the Wilson loop must be placed in the subspace $\varphi
=0$ 
\begin{equation}
W[c(s)]=\int_{x\mid _{\partial M}=c;\varphi \mid _{\partial M}=0}DxD\varphi
e^{-S}
\end{equation}

In other words the open string flies in the $(x,\varphi )$ -space keeping
its feet (its ends) on the ground $(\varphi =0)$. The condition of conformal
invariance, as we will see, induces in this case a contour-dependent $B_{\mu
\nu }$ -field.

In the case of closed strings the general ansatz compatible with the
symmetries of the problem is given by 
\begin{equation}
S=\int d^{2}\xi [(\partial \varphi )^{2}+a^{2}(\varphi )(\partial
x)^{2}+\Phi (\varphi )R\surd g]
\end{equation}

Here $\Phi (\varphi )$ is the dilaton field, $R$ is the curvature of the
world sheet, and, most importantly $a^{2}(\varphi )$ is the running string
tension. The zigzag symmetry requires 
\begin{equation}
a^{2}(0)=0
\end{equation}

Let us discuss this requirement. It is well known that the non-critical
strings with $c<1$ are described by the above effective action (with the
tachyon added) with $a(\varphi )=1$ . At $c=1$ the tachyon becomes
marginally stable and simultaneously the second solution, describing a 2d
black hole appears (see [10] for a review). This second solution has the
variable string tension, satisfying (22). As we move beyond $c=1$ the first
branch with constant tension becomes unstable. We will present now arguments
in favor of the conjecture that as we go beyond $c=1$ the second branch with
the running string tension really describes the string dynamics. It will not
be ruled out however that the stabilization will require the world sheet
supersymmetry (see below).

The exact effective action depending on the background fields $G,B,\Phi $
has the form 
\begin{equation}
\Gamma =\int d\varphi d^{n}x[\frac{1}{3}(c(G,B)-26)-(\nabla \Phi
)^{2}]e^{\Phi }
\end{equation}
Here $c(G,B)\geq 0$ is the central charge of the $\sigma $ -model on
the flat world sheet describing the corresponding target space. It is likely
that the higher derivatives of the dilaton $\Phi $ do not appear in this
action (this can be proved at least for the ansatz (21); see also the
discussion of related matters and references in [10]).
As a side remark, let
us notice that in string theory the only source of the notorious
non-positivity of the Euclidean action comes from the dilaton. This may have
cosmological consequences.

The central charge has the well known perturbative expansion 
\begin{equation}
c=n+1-\frac{1}{2\pi }R+...
\end{equation}
where $R$ is a scalar curvature. At the large positive $R$ the central
charge is not becoming negative (since the formula (24) is not applicable in
this region) but instead tends to zero. This happens because the $\sigma $
-model with positive curvature develops a mass gap. At the large negative $R$
the central charge presumably goes to $+\infty $ (except in the case when
the target space is compact, like a fundamental domain of some discrete
group; in this case the theory has a fixed point).

We are interested in the solution for the ansatz (21). Its substitution in
(23) gives 
\begin{equation}
\Gamma =\int d\varphi e^{\chi }[\frac{1}{3}(n-25)-\stackrel{.}{\chi }^{2}+%
\frac{1}{3}c(\stackrel{.}{\rho },\stackrel{..}{\rho },...)]
\end{equation}

where $\rho =\log (a),\chi =\Phi +n\log (a).$ In the one loop approximation
we have $c\approx \stackrel{.}{\rho }^{2}$ and this action as well as the
corresponding solutions are well known. In this case, as was found in many
works (reviewed in [10]) the asymptotic behavior of the string tension is
given by 
\begin{equation}
a(\varphi )\sim \varphi ^{\frac{1}{\surd n}}
\end{equation}

Unfortunately the curvature in this limit becomes large and negative. That
invalidates the one loop approximation. What really defines the asymptotic
behaviour of the string tension $a^{2}(\varphi )$ is the behaviour of the
central charge at very large negative curvatures. This is presently not
known. It is likely that the needed asymptotic can be found, but presently
we have to take the existence of the running tension as an assumption. We
also have to leave open the choice between the solutions with the constant
negative curvatures in which the tensionless point is located at the
infinite distance in $\varphi $ and the more likely case when this distance
is finite.

Even with this limited knowledge we can analyze the consequences of the
running tension. One possibility is to compare directly the operators of the
gauge theory like $TrF_{\mu \nu }^{2}$ with the closed string vertex
operators. The simplest type of the latter is given by 
\begin{equation}
V(p)=\int d^{2}\xi \Psi _{p}(\varphi (\xi ))e^{ipx(\xi )}
\end{equation}
with the wave functions $\Psi _{p}(\varphi )$ determined again from the
conditions of conformal invariance. In the one loop approximation it is easy
to calculate all these quantities and to see that the correlation function
are power-like in the momentum space. Unfortunately this approximation is
inadequate and we still can't compare these correlation functions with their
values predicted by asymptotic freedom.

Let us discuss other approaches to the problem. The Wilson loop which we
discussed so far is a somewhat difficult object for the non-critical string
theory since the corresponding functional integral includes (in the
conformal gauge) averaging over the reparametrizations of the boundary. This
averaging presents an interesting but unsolved problem, which we hope to
discuss elsewhere. Presently we will introduce two alternative descriptions
to circumvent this problem and to simplify the comparison between strings
and gauge theories.

The first alternative is to replace the contours with the D0 branes. Let us
notice that confining strings, among other things, must pass what we will
call ''the zero-brane test''. Let us take first the straight line as a
contour. It is parametrized as follows 
\begin{equation}
c^{1}(s)=s;c^{k}(s)=0;k=2,...n
\end{equation}

Compare the functional integral for this contour with the one characterized
by the D0-brane boundary conditions 
\begin{equation}
c^{k}\mid _{\partial M}=0;c^{1}\mid _{\partial M}=free
\end{equation}

The difference between these conditions is that in the first case the
surface is attached to the straight line without folds and voids, while in
the second case folds and voids are allowed and actually present. In the
standard non-critical string the two cases are different. In the case of the
confining string the zigzag symmetry makes contours and D0-branes
equivalent(and that is what we called the ''zero brane test''). Let us now
consider a small deformation of the D0-brane 
\begin{equation}
c^{1}\mid _{\partial M}=free;c^{k}\mid _{\partial M}=\zeta ^{k}(c^{1})
\end{equation}
and examine the $\zeta $ -dependence in the string and gauge theories. In
the gauge theory it is easy to calculate this dependence in the case of
small $\zeta $ which are non-zero at the small range of $c^{1}$ (they
represent a small bump on the straight line). In this case the answer is
dominated by the one gluon exchange 
with the renormalized charge and is given by 
\begin{equation}
W\sim \sum_{p}\frac{\mid p\mid ^{3}}{\log \mid \frac{p}{m}\mid }\widetilde{%
\zeta }^{k}(p)\widetilde{\zeta }^{k}(-p)
\end{equation}
where we passed to the momentum representation in $c^{1}$ with $p$ being the
momentum, and introduced the gauge theory scale $m$ . The formula is valid
if $p\gg m$ and $m\zeta \ll 1.$

At the same time in string theory the D0-brane contribution is given by the
expectation value [11] 
\begin{equation}
W[\zeta ]=\langle \exp i\oint_{\partial M}\zeta ^{k}(c^{1})\frac{\partial
x^{k}}{\partial n}ds\rangle
\end{equation}
where the integral is taken along the boundary of the disk and $n$ is the
direction of the normal.

In the standard string $W[\zeta ]$ is described in the low energy limit by
the Born-Infeld action. In the confining string its calculation is still an
open problem. It requires the evaluation of the two-point function of the
vertex operators 
\begin{equation}
V_{k}(p)=\oint \frac{\partial x^{k}}{\partial n}e^{ipx^{1}}ds
\end{equation}

and the formula (31) tells us what the answer should be.

The second alternative is to replace the contour dependence of the Wilson
loop by the dependence on the external field. Let us consider the following
functional 
\begin{equation}
Z[B_{\mu }(x)]=\sum_{(C)}TrPe^{\oint B(x)dx}W(C)
\end{equation}

The external field $B_{\mu }(x)$ may be arbitrary. In terms of string theory
we are introducing a boundary vector field , which is a standard procedure.
In the gauge theory this object is somewhat unusual. In particular, much
care is needed to define the summation over contours in the formula (34). We
try to define it as follows 
\begin{equation}
\sum_{(C)}(...)=\lim_{L\rightarrow \infty }N(L)\int Dx(s)\exp [-\frac{1}{L}%
\int_{0}^{1}ds(\frac{dx}{ds})^{2}](...)
\end{equation}

Typically one needs an infinite factor $N(L)$ to make it well defined. The
whole operation seems much better defined if one introduces supersymmetry on
the world line by considering $x^{\mu }=x^{\mu }(s,\vartheta )=x^{\mu
}(s)+\vartheta \psi ^{\mu }(s)$. In order to have a non-trivial object we
must take $x$ to be periodic while $\psi $ to be antiperiodic on the world
line. We also have to add the term $\int \psi \frac{d\psi }{ds}ds$ in the
exponential in (35).The resulting object is equal to 
\begin{equation}
Z[B]=\lim_{L\rightarrow \infty }\langle \exp (-LD^{2}(A+B)\rangle _{A}
\end{equation}
where $D(A)$ is the Dirac operator in the gauge field $A$. It is interesting
to notice that the Laplace transform of this quantity is the spectral
density of the Dirac operator near zero energy. It has been conjectured [12]
that it can be described by a certain matrix model. This fact may have some
relation to what follows.

We will now introduce a conjecture concerning the equation satisfied by Z.
It seems likely that the conjecture works only for the supersymmetrized
version since we have bad divergencies otherwise. Let us use the following
identity 
\begin{equation}
\sum_{(C)}\widehat{L}(TrPe^{\oint Bdx}W(C))=\int dxa(\Phi _{\mu \nu }(x))
\end{equation}

Here $\Phi $ is the field strength for the gauge field $B$ , the local
function $a(\Phi )$ represents a contribution of the collapsed
loops; this
contribution is not necessarily analytic in $\Phi $. This identity is
conjectural, because the subtle limiting procedures involved in the
functional integration by parts are still to be justified and possible
anomalies to be analyzed. If we trust these naive transformations, this
relation results in the following equation for $Z[B]$%
\begin{equation}
\int dxTr(\nabla _{\mu }\Phi _{\mu \nu }(B)\frac{\delta Z}{\delta B_{\nu }}%
)=g_{0}^{2}\int dxTr(\frac{\delta Z}{\delta B_{\mu }})^{2}+\int dxa(\Phi )
\end{equation}

Somewhat analogous equations have been discussed in [13, 14]. The interest
of this equation is that it has the form of the general string equation for
the open string partition function (7). The only difference between the two
is that in the standard string theory we expect an infinite number of the
intermediate states saturating the pinched disk, while here we have only the
vector state. As was explained above this may be a consequence of the zigzag
symmetry. Unfortunately we can't make the discussion more quantitative until
the explicit string background with the running tension is found. The only
comment to be added is that the gauge theory favors the Wilson super-loops.
This might be a hint that we should be looking for solutions of the string
theory with the supersymmetry on the world sheet (but not in space-time).
Indeed such theories seem to be better defined and may be more suited for
our purposes. However a deeper insight is needed to make a final choice.

Let us now discuss another aspect of our subject. As was explained in [8],
the role of the random surfaces in confining strings can be viewed as the
enforcement of the Bianchi identity on the field strength. Take an abelian
theory first. In this case one has to sum over all possible surfaces
independent of their topology. This summation selects very special
configurations of the field $B_{\mu \nu }.$ Indeed, there is a large class
of fields for which the exponential in (13) does not depend on the surface
at all, thus leading to the divergence of the sum. This is the case when $%
B_{\mu \nu }$ is an abelian field strength, corrected by the quantized
magnetic monopoles 
\begin{equation}
dB=0(mod\ Z)
\end{equation}

It was conjectured in [8] that the summation over surfaces furnishes a $%
\delta $ -function concentrated on those monopoles, and thus reproduces the
old solution of the abelian quark confinement [15]. In other words we
conjectured that in the standard first order formalism of gauge theory with
the action given by 
\begin{equation}
S\sim \int F^{2}+\int B_{n-3}\wedge dF
\end{equation}
where the $n-3$ -form $B$ is a Lagrange multiplier we can replace the
integration over all possible fields $B$ by the integration over 2-surfaces $%
S$ on which this field is concentrated 
\begin{equation}
\ast dB_{n-3}=\int_{S}\delta (x-y)d\sigma _{2}(y)
\end{equation}

The surfaces which appear here are precisely the world surfaces of electric
strings. It must be remembered that in the abelian case one must sum over
the surfaces with arbitrary topology and arbitrary number of the
disconnected components. It is easy to define this summation on a lattice
[8]. Of course in the abelian case the string representation is rather
impractical. Its only role is to motivate the non-abelian string
representation in which case the string coupling constant is $\sim N^{-1}$
and higher topologies are suppressed.

Can the above formulae be generalized to the non-abelian case? Let us
describe a path which may lead to this goal. It is based on Kirillov's
formula for characters (geometric quantization)(see [16] for a recent
review) 
\begin{equation}
TrPe^{\oint Adx}=\int Dg(s)\exp [\int_{0}^{1}ds\frac{dx^{\mu }}{ds}%
Tr(KA_{\mu }^{g})]
\end{equation}

where 
\begin{equation}
A^{g}\frac{dx}{ds}=g^{-1}A\frac{dx}{ds}g+g^{-1}\frac{dg}{ds}
\end{equation}

and $K$ is an element of the Cartan algebra defined by the weight $\mu $ of
the representation in question, $K=(\mu H)$. This element defines an orbit
associated with the representation in question. It is well known that for
the constant $A$ this functional integral can be first localized on the
constant matrices $g$ and then either localized again or directly calculated
by the use of the Itzykson - Zuber formula. The result is the Weyl formula
for characters.

We are looking for the two-dimensional generalization of this construction.
It should provide us with an expression containing a non-abelian 2-form
field $B_{\mu \nu }$ and a random surface $S$ in such a way that the sum
over surfaces is peaked at the 2-forms which are the Yang-Mills field
strengths i.e. expressible through some vector potential. A possible
expression compatible with the zigzag symmetry is given by 
\begin{equation}
S\sim \int d^{2}\xi \epsilon ^{ab}\partial _{a}x^{\mu }\partial _{b}x^{\nu
}Tr(Kg^{-1}B_{\mu \nu }(x(\xi ))g(\xi ))+\Phi [g]
\end{equation}

where $\Phi [g]$ is a 2d action for the chiral field $g(\xi )$ which might
contain the Kirillov 2-form and the WZNW topological term. It is easy to
adjust them so that the Yang-Mills expression for $B_{\mu \nu }$ extremizes
the action. However I have not found the action which picks up the
Yang-Mills configurations only. This is an interesting subject which has
obvious connections with the non-abelian Stokes theorem [17] on one hand and
with the theory of non-abelian 2-forms on the other.

However the above puzzle has little to do with the solution of the loop
equations. The chiral field $g(\xi )$ which represents the non-abelian
electric flux along the string must acquire a finite correlation length in
the confining phase. Its only role must be in providing correct topological
weights to the electric random surfaces. This is the consequence of the fact
that in the confining phase the only memory of the group-theoretic structure
is hidden in the t'Hooft factor $N^{\chi }$ where $\chi $ is the Euler
character of the random surface. Perhaps we can interpret the $\varphi $
-dependent field $B_{\mu \nu }$ of the first part of this paper as a density
of the eigenvalues for the non-abelian field strength. At the present stage
of development it is impossible to make these statements more precise.

Finally let us discuss a possible scenario in higher dimensions. In this
case the Yang -Mills theory is not confining in the weak coupling phase and
must have a phase transition leading to confinement. The nature of this
transition is not known. If we conjecture that it is of the second kind, we
can consider a confining continuous theory in the vicinity of the critical
point. In particular there will be confining strings in the 10d SYM theory.
They well may belong to the same universality class as the type II critical
strings and in particular contain gravitons made of the gauge fields. Let us
stress that the usual objection against the SYM theory in 10d based on its
anomalies is not valid since the structure of the massless states completely
changes in the process of the phase transition. Notice also that in contrast
with the M(atrix) theory we are considering the standard and well defined
large N limit.

We see that better understanding of confinement may also shed light on the
Planck physics.

This work was partially supported by the National Science Foundation under
contract PHYS-90-21984.

\pagebreak

REFERENCES

[1] A. Polyakov Phys. Lett.82B(1979) 247

[2] Yu. Makeenko, A. Migdal Nucl. Phys. B188 (1981) 269

[3] A. Polyakov Nucl. Phys. B164 (1980) 171

[4] V. Kazakov Mod. Phys. Lett A 4,22 (1989) 2125

[5] A. Polyakov Proceedings of the Les Houches School (1993)

[6] A. Polyakov Proceedings of the Les Houches School (1988)

[7] A. Polyakov ''Gauge Fields and Strings'' Harwood Academic Publishers
(1987)

[8] A. Polyakov Nucl Phys B486 (1997) 23

[9] A. Migdal Nucl Phys B265 (1986) 594

[10] A. Tseytlin Class Quantum Grav 12 (1995) 2365

[11] J. Polchinski Preprint HEP-TH /961105

[12] J. Verbaarschot Preprint HEP-TH /9705029

[13] A. Migdal Nucl Phys Proc Suppl 41(1995) 151

[14] H. Verlinde Preprint HEP-TH /9705029

[15] A. Polyakov Phys Lett 59B(1975) 80

[16] R. Jackiw Preprint HEP-TH /9604040

[17] M. Diakonov V. Petrov Preprint HEP-TH /9606104

\end{document}